          [2001/04/25 1.1 (PWD)]
\documentclass[letter,twocolumn]{esapub} % European paper

\usepackage{times}
\usepackage{natbib}
\usepackage{graphicx}
\usepackage{times}

\title{Solar model with CNO revised abundances}
\author{J. Montalb\'an}
\author{A. Miglio}
\author{A. Noels}
\author{N. Grevesse}
\affil{Institut d'Astrophysique et G\'eophysique de l'Universit\'e de Li\`ege, Belgium}
\author{M.P. Di Mauro}
\affil{INAF, Osservatorio Astronomico di Catania, Italy}

\begin{document}

\keywords{Sun: abundances, modelling,  interior}

\maketitle

\begin{abstract}
Recent  three-dimensional, NLTE analyses of the solar spectrum have shown a significant reduction in 
the C, N, O and Ne abundances leading to a Z/X ratio of the order of 0.0177. We have computed solar models with
 this new mixture in the OPAL opacity tables. The present He abundance we find seems rather consistent 
 with the helioseismic value. However, the convective envelope is
  too shallow, and diffusion, even if it reduces the discrepancy, is not able to give the current value.
   We present some numerical experiments consisting in changing the diffusion 
velocities and/or the value of opacity at the base of the convective envelope.

\end{abstract}
\begin{table*}
  \begin{center}
    \caption{Schematic description of the calibrated models according to their chemical composition,
 opacity, diffusion. The last two columns give the bottom of the convective zone and the He abundance 
 at the surface.}\vspace{1em}
    \renewcommand{\arraystretch}{1.2}
    \begin{tabular}[h]{lccccc}
      \hline
      Model  &	Mixture &	Opacity &	Diffusion &	$R_{\rm cz}$ &	$Y{\rm s}$ \\
      \hline
      S1 &	GN93              &	OPAL            &	T (Thoul et al. 94) &	0.714 &	0.246\\
      S2 &	Asplund et al. 04 &	OPAL            &	T                   &	0.727 &	0.243\\
      O1 &	Asplund et al. 04 &	OPAL+Seaton     &	T                   &	0.723 &	0.248\\
      O2 &	Asplund et al. 04 &	OPAL+Bahcall 1  & 	T                   &	0.718 &	0.249\\
      D1 &	Asplund et al. 04 &	OPAL            &	T $\times$ 2        &	0.714 &	0.226\\
      D2 &	Asplund et al. 04 &	OPAL+Seaton     &	T $\times$ 1.5      &	0.717 &	0.239\\
      D3 &	Asplund et al. 04 &	OPAL+Bahcall 2  & 	T $\times$ 1.5      &	0.715 &	0.239\\
      D4 &	Z/X = (Z/X)$_{04} \times 1.1$ &	OPAL    &	T $\times$ 1.5	    &   0.717 &	0.241\\
      \hline
    \end{tabular}  
    \label{tab:table}
  \end{center}
\end{table*}

\section{Modelling the Sun}
\begin{figure*}
\resizebox{\hsize}{!}{\includegraphics{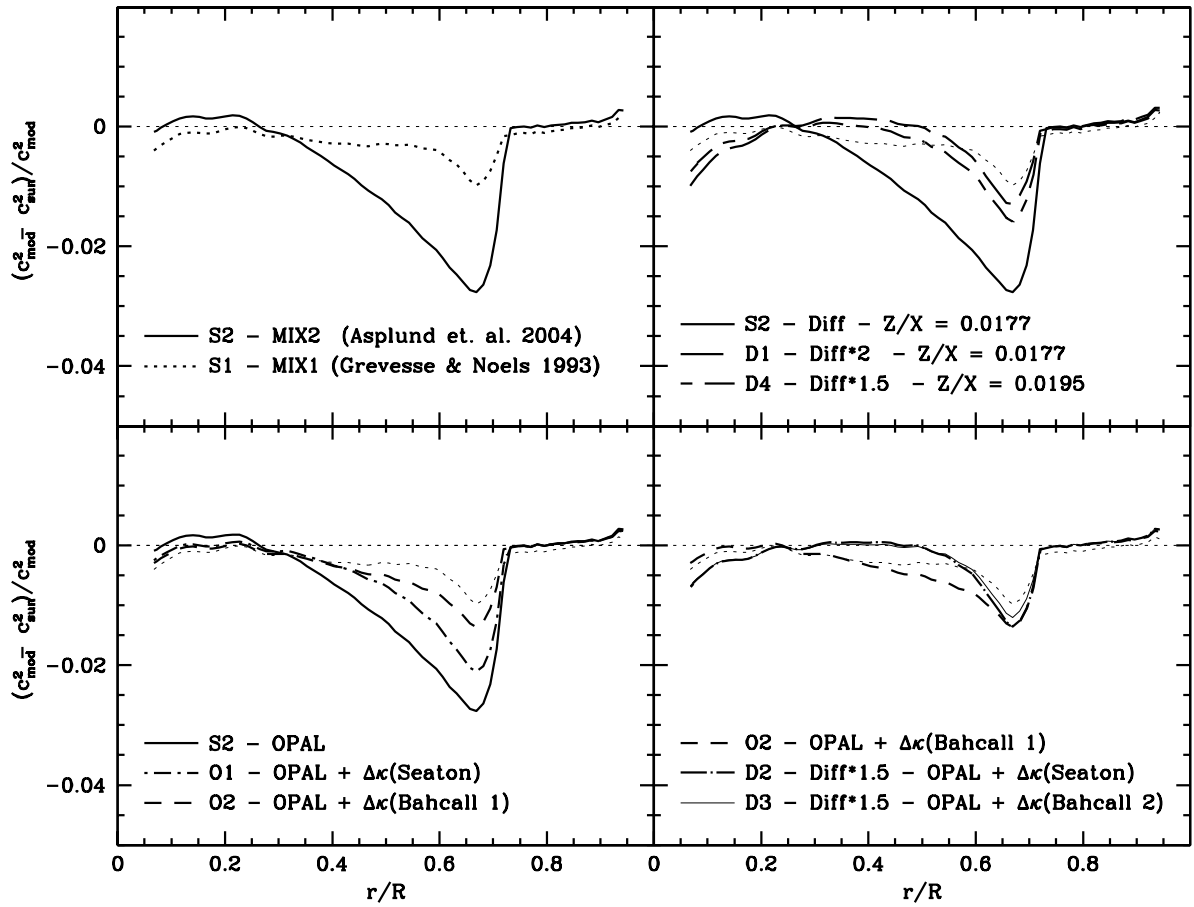}}
\caption{Effects of changes in chemical composition, opacity and diffusion velocities on the error
 in the squared sound speed.}
\end{figure*}

The whole set of models presented here has been produced using the stellar structure and evolution code
 CLES (Code Li\'egeois d'Evolution  Stellaire):
OPAL01 EOS; OPAL96 opacity tables plus  Alexander \& Ferguson (1994) at $T < 6000$~K;
 nuclear reaction rates 
from \citet{caughlan88}; MLT convection treatment; microscopic diffusion of all the elements using 
the subroutine by \citet{thoul94};  atmospheric boundary conditions given by Kurucz (1998) 
at $T=T_{\rm eff}$.
We have used the revised O (Asplund et~al. 2004a) C, N and Ne abundances (Asplund et~al. 2004b) (MIX2).
 EOS and opacity tables have been constructed with this mixture. The calibrated model (model S2, see table 1) is very 
 different from the one (model S1) obtained with the \citet{grevesse93} mixture (MIX1) with regards
  to the distribution of the sound speed as can be seen in Fig.~1a.

  We have tried different numerical experiments in order to reduce this discrepancy:
\begin{itemize}
\item The OPAL opacity near the base of the convective zone has been increased in three different ways. 
In models labeled Seaton (model O1),  the differences between the revised OP and OPAL opacities 
\citep{seaton04} near 2.10$^6$~K are taken into account (Fig.  2, solid line). Models labeled 
Bahcall 1 and Bahcall 2 are constructed according to the suggestion from \citet{bahcall04}\footnote{At the time
of this meeting, the authors kindly warned us about the fact that in an updated version of that paper the value 7\% has 
been corrected and substituted by 21\%.} (Fig. 2, 
long dashed-dashed and dashed lines). Figure 1b shows that the best agreement is obtained with Bahcall 1
 (model O2)
 which means an increase of 14\% in the opacity at the base of the convective envelope. 
 This model is still far from the S1 model (dotted line in Fig. 1b).
\item The diffusion velocities have been increased by factors 1.5 and 2. Figure 1c shows that a better 
sound speed distribution can be reached either by multiplying by a factor 2 the diffusion velocities
 (model D1) or by multiplying them by 1.5 and increasing the Z/X ratio by 10\% (model D4) which is 
 the precision range for this ratio for the Sun.
\item Figures 1b and 1c suggest that the best agreement is reached by increasing both the opacity at 
the base of the convective envelope and the diffusion velocities. Figure 1d shows the effect of an 
increase of $\sim 7$\% in the opacity (solid line and dashed line in Fig.2) and an increase in the diffusion 
velocities by 50\% (models D2 and D3). Both results are much better than those obtained by
 only increasing the opacity by 14\%.          
\end{itemize}

We have checked that the new $^{14}{\rm N}(p,\gamma)^{15}{\rm O}$ astrophysical S-factor
 \citep{formicola04} has no effect on the solar calibration.

\begin{figure}[h]
\resizebox{\hsize}{!}{\includegraphics{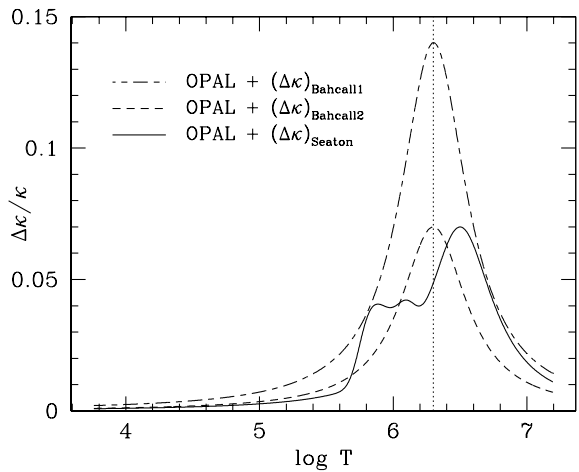}}
\caption{Increases of the opacity (in \%) used in our computations.}
\end{figure}

\section{Conclusion}

The numerical experiments presented here suggest that, in order to reduce the discrepancy between
 the squared sound speed in the Sun \citep{basu00} and that of a theoretical model with the revised 
 C, N, O and Ne abundances, a moderate increase of the opacity (7\%) at the base of the convective 
 envelope should be accompanied by an increase in the diffusion velocities (50\%) (models D2 and D3). 
 Moreover, in these models, the depth of the convective zone is in agreement with the heliosismic value
  (0.713, Christensen-Dalsgaard et~al. 1991) while the surface He abundance is only very slightly smaller 
(0.248, Richard et~al. 1998). These results are equivalent to those obtained by Basu \& Antia (2004) who 
multiplied  the diffusion coefficient by a factor 1.65, and increased the O abundance with respect to
that one given by \citet{asplun04a} (that is, [O/H]=8.80 instead of 8.66).

\section*{Acknowledgments}

J.M., A.M. and A.N. acknowledge the support of the ESA--PRODEX  contract  15448/01/NL/SFe(IC)-C90135.

\begin{small}

\end{small}

\end{document}